\documentclass[twocolumn,aps,pra,showpacs]{revtex4}

\usepackage{amsmath}
\usepackage{graphicx}

\begin{document}
\title{Experimental noise-resistant Bell-inequality violations for
polarization-entangled photons}
\author{Fabio A. Bovino}
\email{fabio.bovino@elsag.it}
\affiliation{Elsag Spa, Via Puccini 2-16154 Genova, Italy}
\author{Giuseppe Castagnoli}
\affiliation{Elsag Spa, Via Puccini 2-16154 Genova, Italy}
\author{Ad\'{a}n Cabello}
\email{adan@us.es}
\affiliation{Departamento de F\'{\i}sica Aplicada II, Universidad de Sevilla, 41012
Sevilla, Spain}
\author{Ant\'{\i}a Lamas-Linares}
\email{antia_lamas@nus.edu.sg}
\affiliation{Quantum Information Technology Lab, Physics Department, National University
of Singapore, 2 Science Drive 3, 117542 Singapore}
\date{\today}


\begin{abstract}
We experimentally demonstrate that violations of Bell's inequalities
for two-photon polarization-entangled states with colored noise are
extremely robust, whereas this is not the case for states with white
noise. Controlling the amount of noise by using the timing
compensation scheme introduced by Kim \emph{et al.}
[Phys. Rev. A \textbf{67}, 010301(R) (2003)], we have
observed violations even for states with very high noise, in
excellent agrement with the predictions of Cabello \textit{et
al.} [Phys. Rev. A \textbf{72}, 052112 (2005)].
\end{abstract}



\maketitle


\section{Introduction}


Entanglement, \textquotedblleft the characteristic trait of
quantum mechanics\textquotedblright~\cite{Schrodinger35}, is
central in Einstein, Podolsky, and Rosen's (EPR's) argument of
incompleteness of quantum mechanics~\cite{EPR35} and in Bell's
proof that quantum mechanics is incompatible with EPR's local
realistic view of the world~\cite{Bell64,CHSH69}. This debate
stimulated the search for sources of entangled states and the
first experiments on the violation of Bell's
inequalities~\cite{FC72}. Nowadays, however, the role of quantum
entanglement is more ubiquitous. Entanglement is considered a
physical resource and a key ingredient for quantum information
processing and quantum computation~\cite{NC00,BEZ00,MPZ00,VW02}.
The challenge now for the development of quantum information
technologies is having reliable and efficient sources to produce,
distribute, and detect entangled states. Although sources of
entanglement have been described and demonstrated in many branches
of physics, so far the most common way to \emph{distribute}
entanglement is by means of pairs of photons. The most reliable
source of entanglement between photons is the spontaneous
parametric down conversion (SPDC)
process~\cite{Klyshko88,KMWZSS95}. The importance of having an
accurate description of the distributed entangled states created
in SPDC processes is therefore clear.

The violation of Bell's inequalities provides a basic tool with which to
detect entanglement~\cite{Terhal98}. In realistic applications, where pure
entangled states become mixed states due to different types of noise,
violations of Bell's inequalities provide a method to characterize the
robustness of the entanglement against noise. For this purpose, different
methods for creating two-photon polarization mixed states have been
proposed, analyzed, and tested~\cite{PAJBK03,PBGWK05,WABGJJKMP05}.

It has been recently pointed out~\cite{CFL05} that a colorless noise model
is not the best choice for describing states produced in type~II SPDC, but
that a more realistic description is given by an alternative one parameter
model where a maximally entangled state is mixed with decoherence terms in a
preferred polarization basis. It turns out to be that this distinction
between colorless and colored noise is crucial when we look for maximal
violations of Bell's inequalities for bipartite entangled states.


In real applications, an example of colored noise could be due to
the first order polarization mode dispersion (PMD) phenomena
because of birefringence in optical fibers. There are different
manifestations of PMD depending on the view taken; in the
frequency domain, one sees, for a fixed input polarization, a
change with frequency $\omega$ of the output polarization; in the
time domain, one observes a mean time delay of a pulse traversing
the fiber which is a function of the polarization of the input
pulse.

Poole and Wagner~\cite{PW86} discovered that there exists special
orthogonal pairs of polarization at the input and the output of
the fiber called the PSPs. Light launched in a PSP does not change
polarization at the output to first order in $\omega$. These PSPs
have group delays $\tau _{g}$, which are the maximum and minimum
mean time delays of the time domain view.

Some insight into the PMD problem can be made simply by
contemplating a piece of polarization-maintaining fiber. Its PSPs
are the polarizations along the principal axes of birefringence of
the fiber. Let us suppose that a polarization entangled state is
launched in this kind of fiber: after the propagation the state
will be affected by colored noise due to the different group
velocity experimented by photons in the birefringent medium.


Violations of Bell's inequalities for two-photon
polarization-entangled states with colored noise are extremely
robust against noise, whereas this is not the case for states with
white noise~\cite{CFL05}. From the experimental point of view, there
exists some evidence supporting this predicted behavior in
experiments with two-qutrit states and low noise~\cite{HLB02}. The
aim of this paper is to experimentally test the predictions
of~\cite{CFL05} for the case of two-qubit states for the full range
of values of the parameter characterizing the amount of noise.


SPDC may be viewed as a three-photon coherent process: a
noncentrosymmetric crystal is illuminated by a pump laser
intense enough to stimulate nonlinear effects. The second order
interaction results in the annihilation of a pump photon and the
creation of two down converted photons, entangled in space-time
or, equivalently, in wave number-frequency~\cite{Rubin96}.
Specifically, in type~II SPDC the incident pump is split in a pair
of orthogonally polarized photons.

Phase-matching by itself does not guarantee entanglement because, if we
restrict our attention to the photons found in the intersections of the
cones made by the extraordinary ($e$) and ordinary ($o$) rays exiting the
crystal, the two down-conversion alternatives $\left\vert
o_{1}\right\rangle\left\vert e_{2}\right\rangle$ and $\left\vert
e_{1}\right\rangle\left\vert o_{2}\right\rangle$ are not indistinguishable.
The indistinguishability is achieved by the use of different kinds of
compensation schemes. These schemes, however, are typically not perfect. For
this reason, while correlations are very strong in the natural basis of the
crystal, the same is not true for the maximally conjugated basis. A
realistic description of the states produced in type~II SPDC is given by a
one parameter model, where a pure state is mixed with decoherence terms in a
preferred polarization basis. These states with colored noise can be
expressed as
\begin{eqnarray}
\rho_{C} & = & p\left\vert \Phi^{+}\right\rangle\left\langle \Phi
^{+}\right\vert+\frac{1-p}{2}\left(\left\vert o_{1}\right\rangle \left\vert
o_{2}\right\rangle\left\langle o_{1}\right\vert \left\langle
o_{2}\right\vert\right. \notag \\
& &\left. + \left\vert e_{1}\right\rangle\left\vert e_{2}\right\rangle
\left\langle e_{1}\right\vert\left\langle e_{2}\right\vert\right),
\label{coloredstates}
\end{eqnarray}
where
\begin{equation}
\left\vert \Phi ^{+}\right\rangle =\frac{1}{\sqrt{2}}\left(\left\vert
o_{1}\right\rangle\left\vert o_{2}\right\rangle +\left\vert
e_{1}\right\rangle\left\vert e_{2}\right\rangle\right).
\end{equation}
The parameter $p$ is the probability of creating the Bell state $\left\vert
\Phi ^{+}\right\rangle$.


If we are interested in maximal violations of the most general
two-party two-output Bell inequality, the
Clauser-Horne-Shimony-Holt (CHSH) inequality~\cite{CHSH69} given
by
\begin{equation}
\left\vert \beta\right\vert \leq 2,
\end{equation}
where the Bell operator is
\begin{equation}
\beta = \langle \hat{A}_{0}\hat{B}_{0}\rangle+ \langle
\hat{A}_{0}\hat{B} _{1}\rangle+ \langle
\hat{A}_{1}\hat{B}_{0}\rangle- \langle
\hat{A}_{1}\hat{B}_{1}\rangle,
\label{beta}
\end{equation}
then, it is sufficient to consider the following local observables:
\begin{eqnarray}
\hat{A}_{0} & = & \sigma_{z},
\label{A0} \\
\hat{A}_{1} & = & \cos\left(2\theta\right) \sigma_{z}+\sin \left(2\theta
\right) \sigma_{x}, \\
\hat{B}_{0} & = & \cos\left(2\phi\right) \sigma_{z}+\sin\left(2\phi \right)
\sigma_{x}, \\
\hat{B}_{1} & = & \cos\left(2\phi -2\theta\right) \sigma_{z}+\sin
\left(2\phi -2\theta\right) \sigma_{x},
\label{B1}
\end{eqnarray}
where $\sigma_{z}$ and $\sigma_{x}$ denote the usual Pauli matrices.

For $\rho_{C}$ states~(\ref{coloredstates}) and the local
observables (\ref{A0})--(\ref{B1}), the Bell operator is
\begin{eqnarray}
\beta\left(p, \theta, \phi\right) & = & \cos\left(2\phi
\right)\left[ \left(1+p\right)\sin^{2}\left(2\theta\right)
+2\cos\left(2\theta\right)
\right] \notag \\
& & +\sin\left(2\phi\right)\left(1+p\right)\sin\left(2\theta
\right)\left[ 1-\cos\left(2\theta\right)\right].
\end{eqnarray}
The maximum value of $\beta(p, \theta, \phi)$ depends on $p$ in a
complex way (for a similar calculation, see~\cite{CFL05}). The
interesting points are that $\rho_{C}$ states violate the CHSH
inequality for \emph{any} $p$, and that the maximum violations
occur for local observables which depend on $ p$~\cite{CFL05}.


\section{Description of the experiment}


\begin{figure}[tbp]
\centerline{\includegraphics[width=7.6cm]{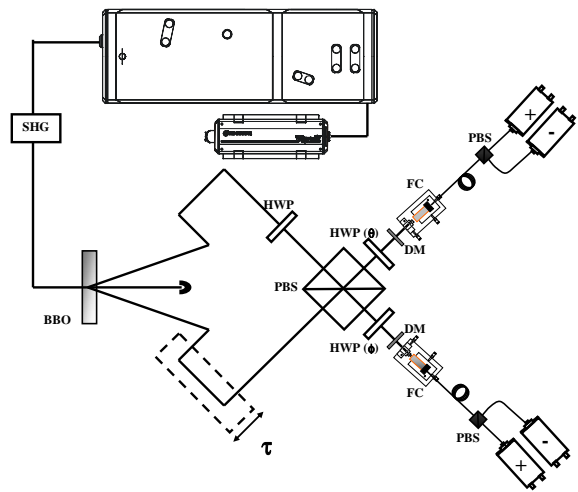}}
\caption{Set-up used in the experiment. A beta-barium borate (BBO)
crystal is pumped by a train of UV ultrafast pulses generated by
the second harmonic of a Ti:sapphire laser producing SPDC photon pairs.
By changing the optical delay introduced by the ``trombone'' in one of
the arms of the interferometer, it is possible to control the
degree of colored noise of the state. After passing through the
interferometer, the photons are coupled by lenses into single-mode
fibers. Dichroic mirrors (DMs) are placed in front of the fiber
couplers (FCs) to reduce stray light due to pump scattering.
Half-wave plates (HWPs) before the fiber couplers, together with
fiber-integrated polarizing beamsplitters (PBSs), project photons
in a specific polarization basis. Photons are then detected by
single photon counters.} \label{Fig1}
\end{figure}


To test this predicted behavior of the colored mixed states in the
laboratory we use the timing compensation scheme introduced by Kim
\emph{et al.}~\cite{KKCGS03,BCDC04,BVCCDS03} described in
Fig.~\ref{Fig1}. In arm~1 of the interferometer, a half wave plate
(HWP) rotates the polarization $90^{ \mathrm{o}}$, so that the
natural emission $\left\vert o_{1}\right\rangle\left\vert
e_{2}\right\rangle$, $\left\vert e_{1}\right\rangle\left\vert
o_{2}\right\rangle$ becomes $\left\vert
o_{1}\right\rangle\left\vert o_{2}\right\rangle$, $\left\vert
e_{1}\right\rangle\left\vert e_{2}\right\rangle$. The state after
the first polarization beamsplitter (PBS) is
\begin{eqnarray}
\left\vert \Psi\right\rangle & = & C\int_{-L}^{0}dz \int
d\nu_{p}\tilde{E} \left(\nu_{p}\right)
e^{i\Lambda_{p}\nu_{p}z}\int d\nu e^{-iD_{G}\nu z}
\notag \\
& & \times \hat{a}_{1o}^{\mathbf{+}}\left(\nu
+\frac{\Omega_{p}+\nu_{p}}{2} \right)
\hat{a}_{2o}^{\mathbf{+}}\left(-\nu +\frac{\Omega_{p}+\nu_{p}}{2}
\right) \notag \\
& & \times e^{i\left(-\nu +\frac{\Omega_{p}+\nu_{p}}{2}\right)
\tau} \notag
\\
& & +\hat{a}_{1e}^{\mathbf{+}}\left(\nu +\frac{\Omega_{p}+\nu_{p}}{2}\right)
\hat{a}_{2e}^{\mathbf{+}}\left(-\nu +\frac{\Omega_{p}+\nu_{p}}{2}\right)
\notag \\
& & \times e^{i\left(\nu +\frac{\Omega_{p}+\nu_{p}}{2}\right)
\tau}\left\vert 0\right\rangle,
\end{eqnarray}
where $\tilde{E}\left(\nu_{p}\right)$ describes the spectral distribution of
the pump field and $\Omega_{p}$ is the central wavelength, and
\begin{eqnarray}
\Lambda_{p} & = &
\frac{1}{u_{p}\left(\Omega_{p}\right)}-\frac{1}{2}\left(
\frac{1}{u_{o}
\left(\Omega_{p}/2\right)}+\frac{1}{u_{e}\left(\Omega_{p}/2
\right)}\right), \\
D_{G} & = & \frac{1}{u_{o}\left(\Omega_{p}/2\right)}-\frac{1}{
u_{e}\left(\Omega_{p}/2\right)},
\end{eqnarray}
where $u_{p}\left(\Omega_{p}\right)$,
$u_{o}\left(\Omega_{p}/2\right)$, and $
u_{e}\left(\Omega_{p}/2\right)$ are, respectively, the group
velocities of the pump, the $o$-photon, and the $e$-photon inside
the $L$-long crystal. The density matrix (obtained by tracing over
the nonpolarization degrees of freedom), can be expressed as
\begin{equation}
\rho = p\left(\tau\right)\left\vert \Phi ^{+}\right\rangle \left\langle
\Phi^{+}\right\vert +\frac{1-p(\tau)}{2}\left(\left\vert
oo\right\rangle\left\langle oo\right\vert + \left\vert
ee\right\rangle\left\langle ee\right\vert \right),
\end{equation}
where
\begin{equation}
p(\tau) = \mathcal{F}\left(\frac{\tau}{D_{G}L}\right)
\left(1-2\left\vert \frac{\tau}{D_{G}L}\right\vert\right)
{e^{-2\sigma_{p}^{2}\left(
\Lambda_{p}\tau/D_{G}\right)^{2}}},
\end{equation}
being
\begin{equation}
\mathcal{F}(x) = \left\{%
\begin{array}{ll}
1 & \mbox{if }\vert x \vert < 1/2 \\
0 & \mbox{otherwise}%
\end{array}%
\right.
\end{equation}
and $\sigma_{p}$ is the bandwidth of the pump laser. Therefore, by
changing the parameter $\tau $, which is related to the optical
delay introduced by the ``trombone'' in arm~2 of the
interferometer, it is possible to control the degree of colored
noise of the state. For $\tau =0$ we ideally obtain the pure state
$\left\vert \Phi ^{+}\right\rangle$.

In the experimental setup (see Fig.~\ref{Fig1}), a 3\,mm long
beta-barium borate crystal, cut for a type~II phase-matching, is
pumped by a train of UV ($\Omega_{p}=410$\,nm) ultrafast (120\,fs)
pulses generated by the second harmonic of a Ti:sapphire laser. SPDC
photon pairs at 820\,nm $\left(\Omega_{p}/2 \right)$ are generated
with an emission angle of $3^{\mathrm{o}}$. After passing through
the interferometer, the photons are coupled by lenses into
single-mode fibers. Coupling efficiency has been optimized by a
proper engineering of the pump and the collecting mode in
experimental conditions. Dichroic mirrors are placed in front of
the fiber couplers to reduce stray light due to pump scattering.
HWPs before the fiber coupler, together with fiber-integrated
polarizing beamsplitters (PBSs), project photons in the
polarization basis $\left\vert s\left(2\alpha\right)\right\rangle
=\cos\left(\alpha \right)\left\vert
o\right\rangle+\sin\left(\alpha\right)\left\vert e\right\rangle$,
$\left\vert s^{\bot}\left(2\alpha\right)\right\rangle = \sin
\left(\alpha\right)\left\vert H\right\rangle -\cos\left(\alpha
\right)\left\vert V\right\rangle$. Photons are detected by single
photon counters (Perkin-Elmer SPCM-AQR-14).

The local observables $\hat{A}_{0}$, $\hat{A}_{1}$, $\hat{B}_{0}$,
and $\hat{B}_{1}$ can be rewritten for the chosen polarization basis $%
\left\{\left\vert s(2\alpha)\right\rangle, \left\vert
s^{\bot}(2\alpha)\right\rangle\right\}$ as
\begin{eqnarray}
\hat{A}_{0,1}(\alpha) & = &\left\vert s(2\alpha)\right\rangle\left\langle
s(2\alpha)\right\vert -\left\vert s^{\bot}(2\alpha)\right\rangle\left\langle
s^{\bot}(2\alpha)\right\vert, \\
\hat{B}_{0,1}(\beta) & = &\left\vert s(2\beta)\right\rangle\left\langle
s(2\beta)\right\vert -\left\vert s^{\bot}(2\beta)\right\rangle\left\langle
s^{\bot}(2\beta)\right\vert,
\end{eqnarray}
and the correlation function $\langle \hat{A}_{0,1}(\alpha)
\hat{B}_{0,1}(\beta)\rangle$ can be expressed in terms of
coincidence detection probabilities $p_{x_{\alpha},
y_{\beta}}(\alpha, \beta)$ as
\begin{eqnarray}
\langle \hat{A}_{0,1}(\alpha) \hat{B}_{0,1}(\beta) \rangle & = &
p_{+_{\alpha}+_{\beta}}(\alpha, \beta)- p_{+_{\alpha}-_{\beta}}(\alpha,
\beta) \notag \\
& & - p_{-_{\alpha}+_{\beta}}(\alpha, \beta)+
p_{-_{\alpha}-_{\beta}}(\alpha, \beta),
\end{eqnarray}
where $x, y = +, -$ are the two outputs of the integrated PBS and
$p_{x_{\alpha}, y_{\beta}}(\alpha, \beta)$ are expressed in terms
of coincident counts,
\begin{eqnarray}
p_{x_{\alpha}, y_{\beta}}(\alpha, \beta) & = &
N_{x_{\alpha},y_{\beta}}(\alpha, \beta)/\left(
N_{+_{\alpha}+_{\beta}}(\alpha, \beta)\right. \notag \\
& & + N_{+_{\alpha}-_{\beta}}(\alpha, \beta)+
N_{-_{\alpha}+_{\beta}}(\alpha, \beta) \notag \\
& & \left. + N_{-_{\alpha}-_{\beta}}(\alpha, \beta)\right),
\end{eqnarray}
where $N_{x_{\alpha}, y_{\beta}}(\alpha, \beta)$ is the number of
coincidences measured by the pair of detectors $x_{\alpha}, y_{\beta}$ in
the polarization basis described above.


\begin{figure}[tbp]
\centerline{\includegraphics[width=9.5cm]{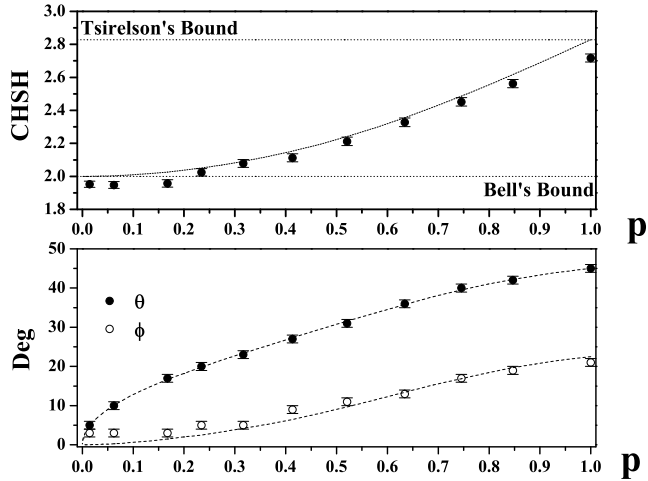}}
\caption{Maximum values of the Bell operator for any set of
measurements for different values of $p$ (above) and the
corresponding local angles giving those maximum values (below). In
both cases dashed lines represent the theoretical predictions of
the model.}
\label{Fig2}
\end{figure}

\begin{figure}[tbp]
\centerline{\includegraphics[width=9.2cm]{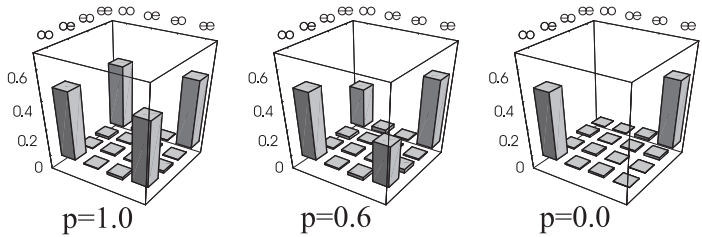}}
\caption{Tomographic reconstructions of the density matrices for
three different values of $p$. The tomographic reconstruction of
the two extremal cases ($p=1$ and $p=0$) and one intermediate case
with a moderate amount of noise ($p=0.6$) independently show that
the colored noise model is a good description of the produced
states.} \label{Fig3}
\end{figure}
\begin{figure}[tbp]
\centerline{\includegraphics[width=8.6cm]{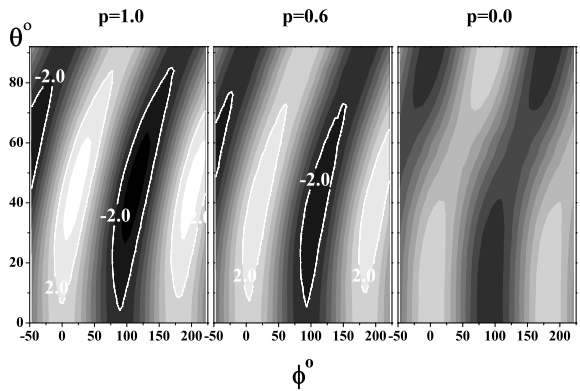}}
\caption{Experimental reconstruction of the Bell operator
$\protect\beta$ for three different values of $p$.} \label{Fig4}
\end{figure}

The experimental results are presented in Figs.~\ref{Fig2}--\ref{Fig4}. In Fig.~\ref{Fig2} we present the
maximum violations of the CHSH inequality [i.e., the maximum
values of the Bell operator $\beta$ given by (\ref{beta})] for
11~values of the parameter $p$ characterizing the amount of noise
and the corresponding local parameters giving those maximum
values. There is a very good qualitative and quantitative
agreement between the experimental data and the theoretical
predictions~\cite{CFL05}. Specifically, we observe violations of
the CHSH inequality even for low values of $p$. Note, however,
that the experimental maximum violations of the CHSH stands
slightly below the theoretical predictions. This effect is due to
a little amount of white noise imputable to experimental
imperfections. This residual white noise is visible in the
tomographic reconstructions of the density matrices for three
values of $p$ ($1.0$, $0.6$, and $0.0$) presented in
Fig.~\ref{Fig3}.


In addition, in Fig.~\ref{Fig4} we present three contour plots representing
the experimental reconstruction of the Bell operator $\beta$ for three
values of $p$ ($1.0$, $0.6$, and $0.0$).


\section{Conclusions}


Summing up, we have performed a range of Bell's inequality tests and
tomographic analysis on two-photon polarization states created by
type~II SPDC processes affected by a controlled colored noise. We
have shown that these states violate the CHSH inequality, even the
very noise ones, in excellent agreement with the theoretical
predictions in~\cite{CFL05}. Our results should have a direct
relevance to the problems of controlling, manipulating and mapping
quantum states for quantum technologies. Specifically, this
description should be useful for optimizing protocols for the
distillation of Bell states~\cite{BBPSSW96} and for quantifying the
security of realistic quantum key distribution schemes based on
pairs of polarization-entangled photons~\cite{JSWWZ00}.


\section*{Acknowledgments}


These experiments were carried out in the Quantum Optics Laboratory
of Elsag Spa, Genova. F.A.B. acknowledge support from EC-FET Project
No.~QAP-2005-015848. A.C. acknowledges support from Projects
No.~FIS2005-07689 and No.~FQM-239. A.L.-L. acknowledges support from
Grant No.~R-144-000-137-112.


\end{document}